\documentclass[%
 reprint,
 amsmath,amssymb,
 aps,
pre,
]{revtex4-2}
\usepackage{float}
\usepackage{siunitx}
\usepackage{color,soul}
\usepackage{amsmath}
\usepackage{bm}
\usepackage{fullpage}
\usepackage{ragged2e}
\usepackage{blindtext}
\usepackage{hyperref}
\usepackage{amsfonts}
\usepackage{graphicx} 
\usepackage{tikz}
\usepackage{bbm}
\usepackage{bbold}
\usepackage{hyperref}
\usepackage{float}
\newtheorem{theorem}{Theorem}

\DeclareMathOperator*{\minimize}{minimize}

\begin{document}
 
\title{The Feasibility of Acoustophoresis Multimodal Control}

\author{Guilherme Perticarari}
\email{guilherme.junqueira\_perticarari@eit.lth.se}
\author{Dongjun Wu}%
\email{dongjun.wu@control.lth.se}
\affiliation{Lund University, Department of Automatic Control}%
\author{Thierry Baasch}
\email{thierry.baasch@bme.lth.se}
\affiliation{Lund University, Department of Biomedical Engineering}%

\begin{abstract}

Actuating the acoustic resonance modes of a microfluidic device containing suspended particles  (e.g., cells) allows for the manipulation of their individual positions. In this work, we investigate how the number of resonance modes $M$ chosen for actuation and the number of particles $P$ affect the probability of success $S$ of manipulation tasks, denoted Acoustophoretic Control Problems (ACPs). Using simulations, we show that the ratio of locally controllable volume to the state-space volume correlates strongly with $S$. This ratio can be efficiently computed from the pressure field geometry as it does not involve solving a control problem, thus opening possibilities for experimental and numerical device optimization routines. Further, we show numerically that in noise-free 1D systems $S \approx 1 - P/M$, and that in noisy 1D and 2D systems $S$ is accurately predicted by Wendel's Theorem. We also show that the relationship between $M$ and $P$ for a given $S$ is approximately linear, suggesting that as long as $P/M$ is constant, $S$ will remain unchanged. We validate this finding by successfully simulating the control of systems with up to $60$ particles with up to $600$ modes.

\end{abstract}

\maketitle

\section{Introduction}\label{sec:intro}

In 1787, Ernst Chladni demonstrated that sand particles scattered over a metal plate can be moved by exciting the plate's resonance modes \cite{ chladni1787entdeckungen}. Since this famous experiment, researchers have tried to exploit this phenomenon, called \textit{acoustophoresis}, for active manipulation of small particles. Recently, it has been experimentally demonstrated that combining multiple vibration modes with feedback control algorithms enables the precise control of up to four particles on centimeter-sized Chladni plates \cite{zhou2016controlling, latifi2019motion, latifi2020model, kopitca2021programmable}. The technology has been further applied for the active control of particles and droplets suspended in acoustically actuated microfluidic chips and capillaries \cite{yang2022harmonic, veikko_model, yiannacou2021controlled, shaglwf}. Feedback controlled acoustophoresis can also be used to control swarms of micro robots \cite{schrage2023ultrasound}. Although this cutting-edge technology has the potential to rival optical tweezers in terms of flexibility and cost, little is known about the theoretical limitations of the number of particles that can be controlled independently. 

To the best of our knowledge, the only statement in the literature so far concerning the systems controllability has found that at least 2P + 1 modes are required to achieve full control of P particles on a Chladni plate \cite{zhou2016controlling}. 
In this work, we use Monte Carlo simulations to show that the success rate of randomly selected multimodal control tasks correlates strongly with the local controllability ratio, a parameter based on the pressure field geometry that is closely tied to Wendel’s theorem, described in Section \ref{sec:results}. If a simple ideal 1D problem is considered, the success rate of a randomly selected control task $S$ using $M$ modes is well approximated by $1 – P/M$. Our findings suggest that to keep the success rate of a (general) manipulation task constant, the ratio between the number of controlled particles P and the number of applied modes M also needs to be held constant. The result is validated by performing numerical pick and place experiments in a system with up to 60 particles using a quadratic programming algorithm and 600 frequencies.
\begin{figure}[htb]
    \centering
    \includegraphics[width=\linewidth]{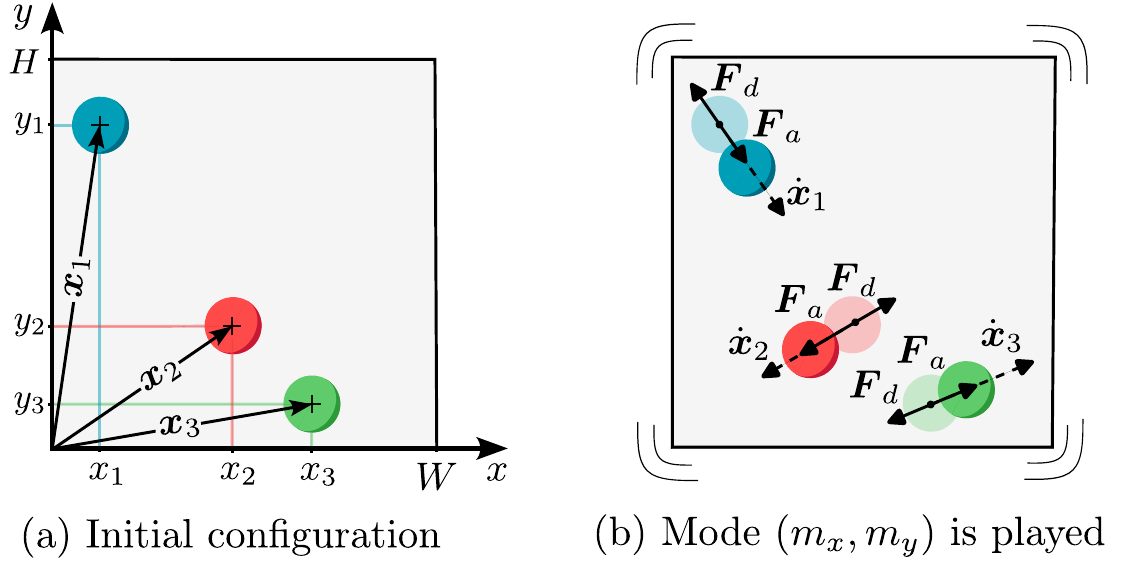}
    \caption{\textbf{(a)} Three particles are introduced into the rectangular device with width $W$ and heigh $H$ at positions $\bm{x}_1=(x_1,y_1)$, $\bm{x}_2=(x_2,y_2)$, and $\bm{x}_3=(x_3,y_3)$. \textbf{(b)} Actuating the mode $(m_x, m_y)$ induces the acoustic radiation force $\boldsymbol{F}_a$, which is balanced by the Stokes drag force $\bm{F}_d$. The effect results in particles moving with velocity $\dot{\boldsymbol{x}}$ in the direction of $\bm{F}_a$.}
    \label{fig:time-step}
\end{figure}
\section{System Dynamics}
\subsection{The Physics of Acoustophoresis}\label{subsec:physics}

The acoustic pressure $p$ inside a rigid walled rectangular device can be modeled by the wave equation \cite{Landau1987Fluid}, formally, 
\begin{equation} \label{eq:wave}
    \nabla^2p(x,y,t)  - \dfrac{1}{c^2} \dfrac{\partial^2}{\partial t^2}p(x,y,t) = 0,
\end{equation}
where $t$ is the time, $c$ is the speed of sound inside the fluid, and $x$ and $y$ denote spatial coordinates measured from the device's rigid walls along its width $W$ and height $H$, respectively. The pressure field's corresponding acoustic velocity field is given by
\begin{equation}
\rho_0 \frac{\partial}{\partial t} \bm{v}(x,y,t) = - \bm{\nabla} p(x,y,t),
\end{equation}
where $\rho_0$ denotes the fluid's unperturbed density.

If the device has width $W$ and height $H$, and $m_x$ and $m_y$ are the system's $x$ and $y$ acoustic mode numbers, then the acoustic pressure is given by
\begin{equation} \label{eq:pressure}
    p(x,y,t) = A \cos \left( m_x\pi \dfrac{x}{W} \right) \cos \left( m_y \pi \dfrac{y}{H}y \right) \cos \left( \omega t \right),
\end{equation}
 where A is the amplitude, and the angular frequency $\omega$ is given by
 \begin{equation}
     \omega = \pi c \sqrt{{\left(\frac{m_x}{W}\right)}^2 + {\left(\frac{m_y}{H}\right)}^2}.
 \end{equation} 
Note that a rigid-walled device implies $\bm{v} \cdot \bm{n}=0$, where $\bm{n}$ is the normal vector to the wall.

A particle introduced in the sound field may scatter the wave and exchange momentum with it. This transfer of momentum results in a force being applied to the particle, causing it to move. For a particle of radius $a$ much smaller than the wave-length of the wave, the induced force $\bm{F}_a$, called \textit{acoustic radiation force}, can be modeled as the negative gradient of the Gor'kov potential $U$ \cite{gor1962forces}, formally,
\begin{equation} \label{eq:force}
    \bm{F}_a = - \bm{\nabla} U.
\end{equation}
where the $U$ is given by 
\begin{equation} \label{eq:u}
    U = V_p \left( \frac{f_1}{2 \rho_0c^2} \langle pp \rangle - \frac{3}{4} f_2 \rho_0 \langle \boldsymbol{v} \cdot \boldsymbol{v} \rangle \right),
\end{equation}
and where $V_p$ is the particle's volume, and $f_1$ and $f_2$ are, respectively, the particle's monopole and dipole scattering coefficients \cite{bruus2012acoustofluidics}. The angled brackets denote the time-average over one actuation period. 

Besides $\bm{F}_a$, the other dominating force acting on the particle is the Stokes drag force, given by
\begin{equation} 
\bm{F}_d = -6 \pi \eta a \dot{\bm{x}},
\end{equation}
where $\eta$ denotes the fluid's dynamic viscosity and $\bm{x} = (x, y)$ is the particle's position.

We assume that the particles are sufficiently far apart from one another so that particle-particle interactions can be neglected. Thus, the only forces acting on each particle are $\bm{F}_a$ and $\bm{F}_d$. Furthermore, \cite{BaaschMylti, C2LC21068A} have shown that inertial effects are typically negligible in acoustophoresis settings. Therefore, Newton's Second Law states that both forces must be balanced, resulting in
\begin{equation} \label{eq:dotx}
\dot{\bm{x}} = \frac{\bm{F}_a}{6 \pi \eta a}.
\end{equation}

Note that by neglecting particle interactions, each particle in the system is governed by  Eq. \ref{eq:dotx} independently, where the right hand side is a function of each of the particle's properties, e.g., $a$, and the resonant mode being induced in the device, i.e., $(m_x,m_y)$. 

Fig. \ref{fig:time-step} illustrates three particles being displaced by $\bm{F}_a$ under the action of resonance mode $m$.

\subsection{State-Space and Mode-Mixing}

The number of resonant modes that an actuator is able to successfully induce is limited in a real device. To model this limitation we choose a number $M$ of modes available to the controller, where $f_{m}^{(p)}(\boldsymbol{x}_p)$ captures the effect of the $m^{\text{th}}$ available mode on particle $p$ via Eq. \ref{eq:dotx} for $m=1,\dots,M$. Thus, if we have a system with $P$ particles where $\boldsymbol{x}_p$ defines particle $p$'s position, then
\begin{equation} \label{eq:pure_mode}
    \text{ mode } m \text{ applied } \implies \left\{
    \begin{aligned}
        \dot{\boldsymbol{x}}_{1} & = f_m^{(1)}(\boldsymbol{x}_1)\\
         & \vdots\\
        \dot{\boldsymbol{x}}_{P} & = f_m^{(P)}(\boldsymbol{x}_P)
        \end{aligned}
    \right.
\end{equation}

Neglecting inertia reduces the order of the equation of motion. The state of a system with $P$ particles is then fully described by the generalized coordinate $\bm{q}$ collecting the positions of all particles 

\begin{equation}
    \bm{q}= \boldsymbol{x}_1 \oplus \dots \oplus \boldsymbol{x}_P,
\end{equation} where $\oplus$ indicates a concatenation of coordinates. 

Moreover, we define the  effect of the $m^{\text{th}}$ mode on $\bm{q}$ as the concatenation of its effect on each $\bm{x}_p$, i.e.,
\begin{equation} \label{eq:f_on_state}
     \dot{\boldsymbol{q}} = f_m(\boldsymbol{q}) = f_m^{(1)}(\boldsymbol{x}_1) \oplus \dots \oplus f_m^{(P)}(\boldsymbol{x}_P)
\end{equation}

The controller is allowed to induce multiple modes simultaneously, i.e., the actuator applies a mixture $\boldsymbol{w} \in \mathbb{R}^M$, with $\bm{w} \geq 0$ and $\sum_i \hat{w}_{i} \leq 1$, of all $M$ modes at once. Note that this mixture is simply a positive combination of all modes.

The effect of playing mixture $\bm{w}$ is given by the weighted sum of the effects of all modes in the mixture if the difference between the actuated frequencies is sufficiently large such that the coupled terms average to zero on the observable time-scale \cite{karlsen2015forces}, i.e., 
\begin{equation} \label{eq:mixture}
        \dot{\boldsymbol{x}}_{p} = \sum_{m=1}^Mw_mf_m^{(p)}(\bm{x}_p), \; \forall p.
\end{equation}
Let us define $\bm{F}_M(\bm{q})$ as a matrix whose $m^{\text{th}}$ column is $f_m(\bm{q})$, then
\begin{equation}
    \dot{\boldsymbol{q}} =\bm{F}_M(\boldsymbol{q})\bm{w}.
    \label{eq:dot_q}
\end{equation}
\subsection{Numerical Simulation}
The equations of motion are numerically integrated using the Euler forward time-stepping algorithm. Given $\bm{w}$ and $\bm{x}_p$ for each particle $p$, Eq. \ref{eq:mixture} is used to calculate $p$'s velocity vector $\dot{\bm{x}}_p$, and then, its position after a time $\Delta t$ is approximated by 
\begin{equation} \label{eq:euler}
    \boldsymbol{x}_p' \approx \boldsymbol{x}_p + \dot{\boldsymbol{x}}_p \Delta t.
\end{equation}
In this work, the dynamics $f_m^{(p)}$ are explicitly calculated using Eq. \ref{eq:dotx} for a device with $W=1600\,$\unit{\micro\meter} and $H=1200\,$\unit{\micro\meter}, a fluid with $\eta=10^{-3}\,$\si{\pascal\second}, $\rho_0=1$\si{\kilo\gram\per\meter\cubed} and  $c=1500\,$\si{\meter\per\second}, particles with $f_1=0.46$ and $f_2=0.03$, chosen to approximate polysteryne, radii sampled uniformly from $2$\si{\micro\meter} to $3$\si{\micro\meter}, and a controller that produces acoustic pressure waves with amplitude $A=10^5\,$\si{\pascal}. Furthermore, for all simulations, we use $\Delta t=5\times10^{-3}$\si{\second}. 

Unless otherwise stated, all experiments will take place in the first quarter of the device, i.e., $[0,W/2] \times [0, H/2]$, because, due to symmetry, under noise-free dynamics particles would not be able to cross the middle horizontal and vertical lines of the device.

\section{Acoustophoretic Control Problem (ACP)}

\begin{figure}[h]
    \centering
    \includegraphics[width=\linewidth]{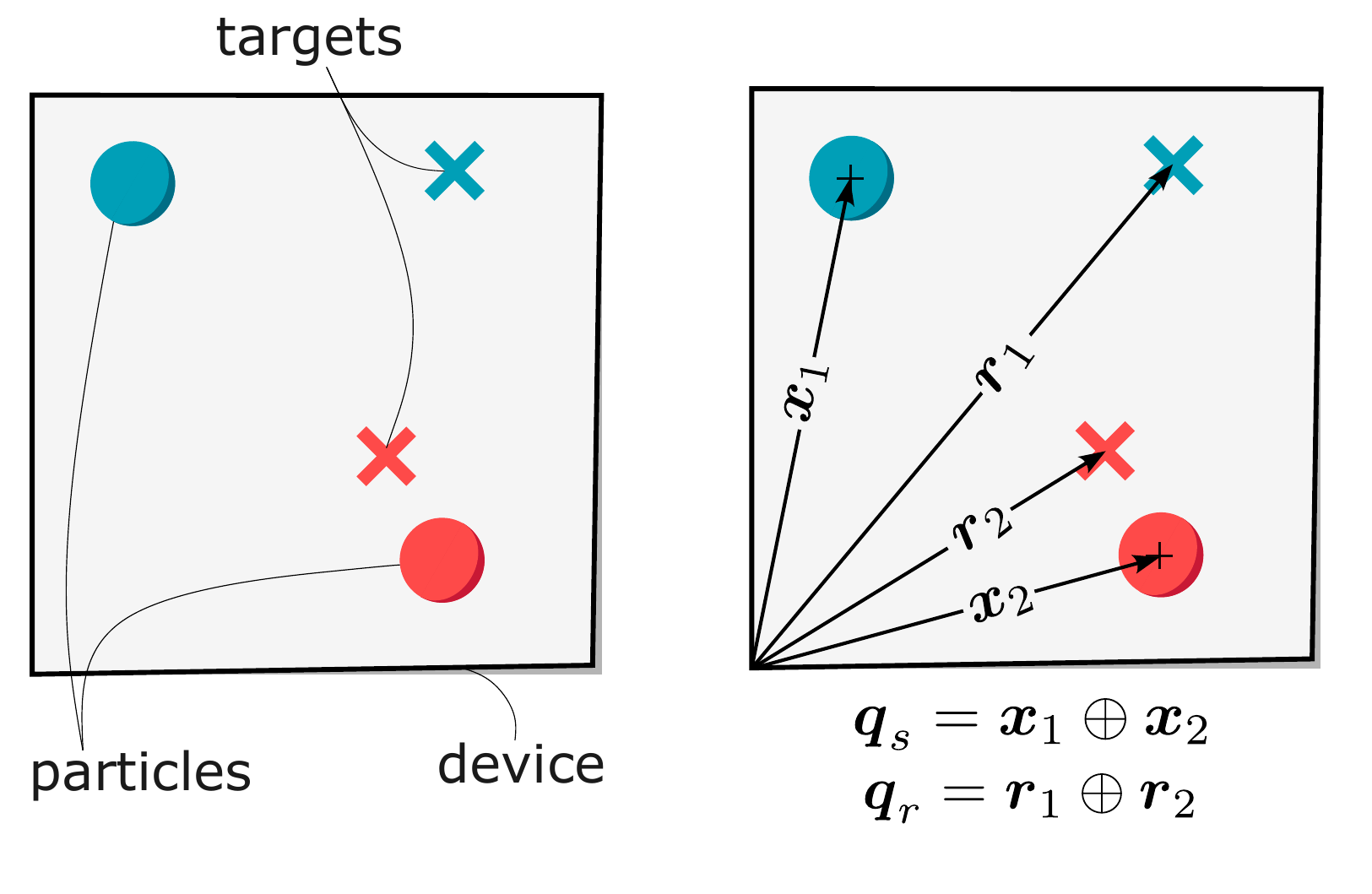}
    \caption{The aim of the Acoustophoretic Control Problem (ACP) is to bring the system from its initial state $\bm{q}_s = \bm{x}_1 \oplus \bm{x}_2$ into its target state $\bm{q}_r = \bm{r}_1 \oplus \bm{r}_2$ using a series of control inputs.}
    \label{fig:system_dynamics}
\end{figure}

A multi-modal particle manipulation task generally involves taking particles from a given state to a target state using an actuator, such as moving them from inlets to outlets \cite{veikko_model}. This control task, which we denote \textit{Acoustophoretic Control Problem (ACP)}, is herein formalized and discussed.

Formally, we define an ACP as a tuple $(\mathcal{S}_P, \bm{F}_M, \bm{q}_s, \bm{q}_r, \theta, T)$, where $\mathcal{S}_P$ is the set of possible states that a system with $P$ particles might be in, and $\bm{q}_s, \bm{q}_r \in \mathcal{S}_P$ denote the system's initial and target states, respectively. A controller applies mode mixture $\bm{w} \in \mathbb{R}^M_+$, with $\sum_i \hat{w}_{i} \leq 1$, resulting in effect $\dot{\bm{q}} = \bm{F}_M(\bm{q})\bm{w}$, see Eq. \ref{eq:dot_q}.  The ACP is considered feasible if the controller is able to take each particle $p$ from its initial position to some $\bm{x}_p$ that is within tolerance $\theta$ of $\boldsymbol{r}_p$, formally, 
\begin{equation}
\underset{p=1,\dots,P}{\max}\left(\|\bm{x}_p - \bm{r}_p\|_2\right) \leq \theta,
\label{eq:distance}
\end{equation}
in under $T$ seconds. Note that Eq. \ref{eq:distance} guarantees that all particles are within tolerance $\theta$ of their targets at the end of a successful experiment.

 By neglecting the inertia,  we consider the states to be solely the spatial configurations of particles (i.e., $\mathcal{S}_P = \mathbb{R}^{2P}$). An example of such an ACP is illustrated in Fig. \ref{fig:system_dynamics}, which depicts a system with 2 particles immersed in 2D acoustic fields and has a degree of freedom of 4. The ACP consists of steering each particle's position $\bm{x}_p$ towards its target $\bm{r}_p$ by actuating the resonance modes. Note that in a 1D problem, we would have $\mathcal{S}_P = \mathbb{R}^P$.

 \subsection{Random Experiments}


Finding a globally optimal solution to an ACP is complex due to the non-linearity of the system dynamics. Therefore, we focus on locally optimal solutions, which currently make up the state-of-the-art solutions for acoustophoretic tasks \cite{yiannacou2021controlled,shaglwf}.

A locally optimal algorithm will use all information concerning the current state $\bm{q}$ to find the mixture $\bm{w}$ that takes the system towards a new state $\bm{q}'$ that is as close as possible to the target $\bm{q}_r$. This optimization problem can be summarized by

\begin{equation} \label{eq:qcpa}
    \begin{aligned}
    \minimize_{\hat{\bm{w}} \in \mathbb{R}^M} \quad & 
        \| \boldsymbol{q}  + \bm{F}_M(\boldsymbol{q}) \Delta t\hat{\bm{w}} - \boldsymbol{q}_r\| _2^2,
        \\
    \textrm{subject to} \quad 
        & \hat{\bm{w}} \geq 0 \text{ and } \sum_i \hat{w}_{i} \leq 1, \\
    \end{aligned}
\end{equation}

We denote the control algorithm that outputs a solution to Eq. \ref{eq:qcpa} as \textbf{P}erfect \textbf{I}nformation \textbf{L}ocal \textbf{O}p\textbf{t}imizer (PILOT), since it uses perfect local information to solve for $\bm{w}$. Variations of PILOT have been used in the literature\cite{shaglwf, zhou2016controlling}, where the predicted displacement $\bm{F}_M(\bm{q})\Delta t$ is estimated at run time.

An ACP's feasibility under PILOT depends on the selected initial and target positions, i.e. it may be feasible to bring a system from configuration $\boldsymbol{q}_s$ to $\boldsymbol{q}_r$, but infeasible to take $\boldsymbol{q}_s'$ to $\boldsymbol{q}_r'$. Therefore, the notion of feasibility is herein generalized to denote the probability of success of an ACP with initial and target positions randomly sampled from a uniform distribution over $\mathcal{S}^P$, and PILOT is used as a control algorithm. We denote the coupling of a random ACP and PILOT as a \textit{random experiment}. Our goal will be to study how the choice for $M$ and $P$ in an ACP influences the probability of success of a random experiment, which we will denote as $S$.

\section{Local Controllability Ratio}\label{sec:defining_R}

Even though defining a random experiment is straightforward, determining its success rate $S$ can be computationally expensive, as it involves running many time-consuming trajectory simulations. Here we show that $S$ is strongly correlated to the system's local controllability ratio $R$, which can be cheaply computed from the pressure field geometry. Therefore, we can gather data for $(M,P,R)$ to study the relationship between $(M,P)$ and $R$, and extrapolate our findings to the success rate of a random experiment $S$. We first define local controllability, and then describe the method with which the local controllability ratio $R$ can be accurately estimated.

\begin{figure}[h]
        \centering
        \includegraphics[width=\linewidth]{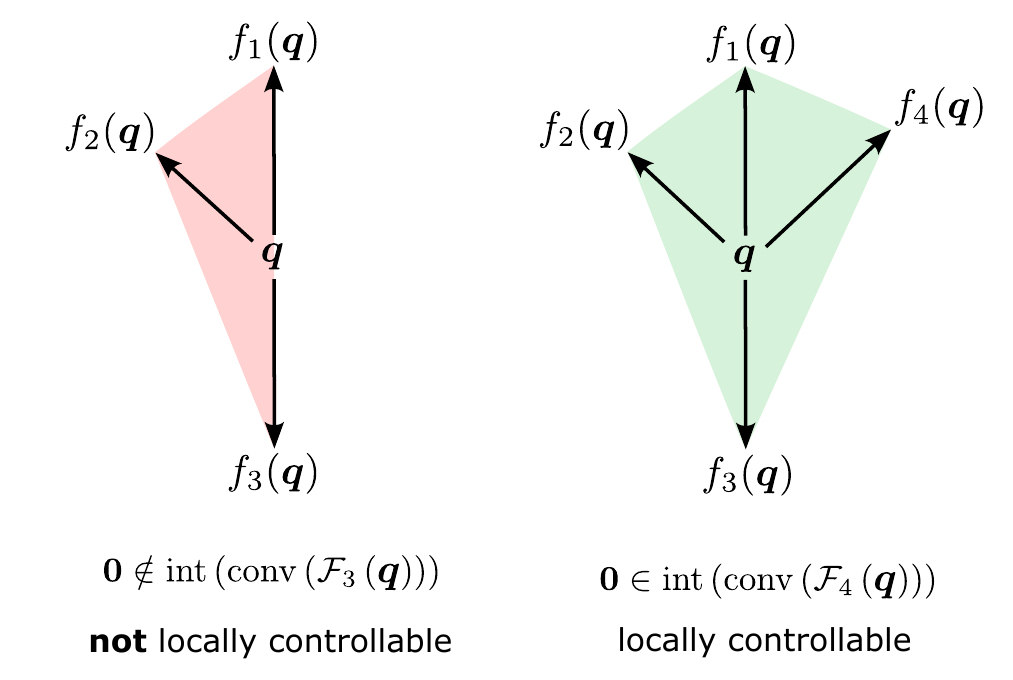}
        \caption{(\textbf{Left}) State $\boldsymbol{q}$ is not locally controllable under $\bm{F}_3(\boldsymbol{q})$ since there are blindspots around it. (\textbf{Right}) State $\boldsymbol{q}$ is locally controllable under $\bm{F}_4(\boldsymbol{q})$.}
        \label{fig:locally}
\end{figure}

A state $\boldsymbol{q}$ is deemed locally controllable if the system can transition from it to any other state in an open neighbourhood around it. This property, thus, simply relies on checking whether $\boldsymbol{0}$ is in the interior of the convex hull of vectors $f_m(\boldsymbol{q})$, i.e. if all directions are included in the set of possible $\dot{\boldsymbol{q}}$, as shown in Fig. \ref{fig:locally}. Note that every mode can only move the system in one direction along its trajectory in the state-space. Therefore, we can control the system only in the space spanned by positive linear combinations of the vectors $f_m(\boldsymbol{q})$. As a consequence, at least three modes are necessary to achieve local controllability of a system of two degrees of freedom. 

A system's local controllability ratio, $R$, is defined as the percentage of its states in the state-space that are locally controllable, i.e. whose velocity vectors $\bm{\dot{q}}$ under each mode $M$ have a positive combination that span the entire state-space. Similarly, $R$ can be understood as the ratio of the locally controllable volume to the state-space volume.

\begin{figure}[h]
    \centering
    \includegraphics[width=\linewidth]{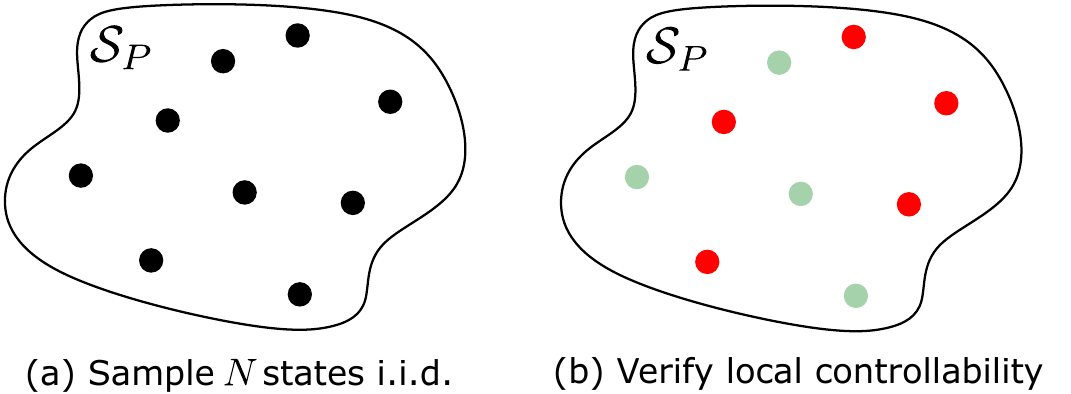}
    \caption{To estimate the local controllability ratio $R$, we sample $N$ states i.i.d. from $\mathcal{S}^P$, verify which are locally controllable (green) and take $R$ as the ratio of locally controllable states to $N$.}
    \label{fig:MonteCarlo}
\end{figure}

We use the Monte-Carlo method as an unbiased estimator for $R$: first we i.i.d. sample a number $N$ of states from the state-space, and use a convex hull inclusion test to verify which are locally controllable, as shown in Fig. \ref{fig:MonteCarlo}. Then, we use the percentage of locally controllable states as an estimate for $R$.
\subsection{$R$ Is a Proxy for $S$}\label{subsec:r}
\begin{figure}[h]
    \centering
    \includegraphics[width=0.9\linewidth]{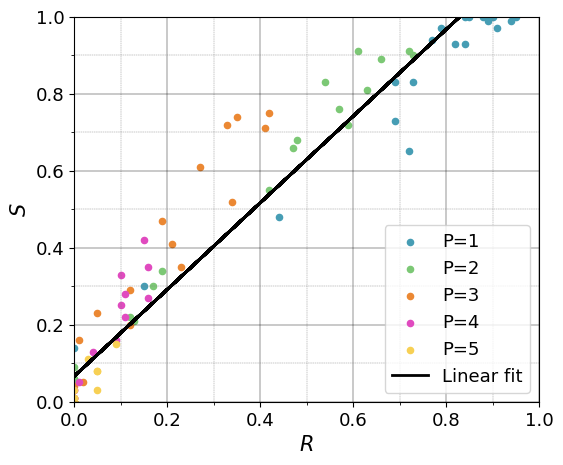}
    \caption{The Local Controllability Ratio, $R$, and the success rate of a random experiment, $S$, are estimated for ACPs with varying number of particles and modes. The data shows a 0.97 correlation between both metrics.}
    \label{fig:corrl_R_success}
\end{figure}

Fig. \ref{fig:corrl_R_success} shows the correlation between $R$ and $S$. The data was created by first generating datasets for $(M,P,R)$ and $(M,P,S)$. These datasets are then joined on $(M,P)$, resulting in the augmented dataset $(M,P,R,S)$, which is then used to estimate the correlation between $R$ and $S$. 

We first generate $100$ $(M,P)$ tuples by varying $P=1,\dots,5$ and $M=1,\dots,20$ and, for each tuple, we sample $N=100$ random states, verify their local controllability and estimate $R$, as done in Fig. \ref{fig:MonteCarlo}. 

Subsequently, for each $(M,P)$, we run $N=100$ random experiments by sampling each particle's initial position uniformly inside the first quarter of the device, and then sampling a target position within a $50$\,\unit{\micro\meter} distance from it. Then, PILOT was used to pick appropriate mode mixtures at each time-step. An experiment was deemed successful whenever the  position-to-target distance of the farthest particle was within a tolerance of 10\,\unit{\micro\meter}  in under 2000 iterations, and the percentage of successful experiments (out of $N$) was taken as our estimate for $S$ for the given $(M,P)$.  

The result, shown in Fig. \ref{fig:corrl_R_success}, shows a Pearson correlation of $0.97$ between $S$ and $R$, which suggests that the much cheaper $R$ can be used as a proxy for $S$.
\section{Results}\label{sec:results}
\begin{figure*}
    \centering
    \includegraphics[width=\textwidth]{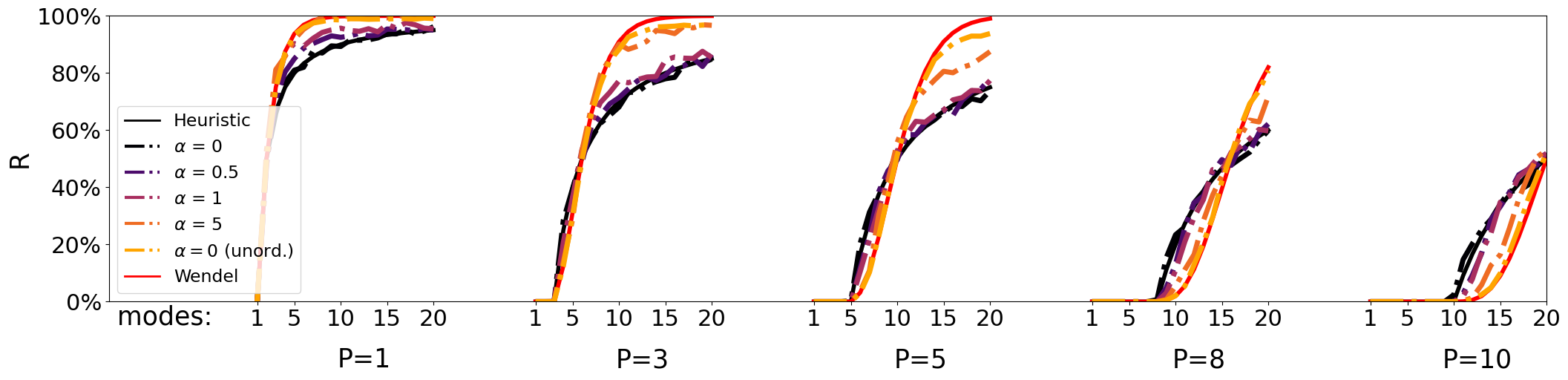}
\includegraphics[width=\textwidth]{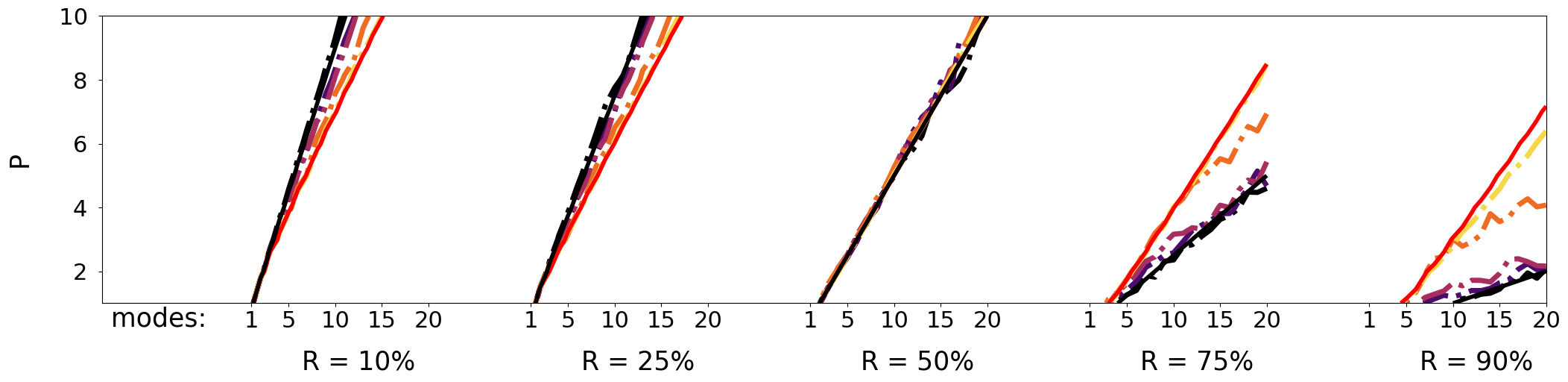}
    \caption{The local controllability ratio $R$ is numerically approximated using Monte Carlo sampling with varying number of particles $P$, modes $M$ and noise levels $\alpha$ in the 1D problem. Additionally, the heuristic $R = 1-P/M$ and the Wendel approximation are shown, respectively, as black and red solid lines. The top plot emphasizes the effect of $P$ and $M$ on $R$, while the bottom one focuses on the linear relationship between $M$ and $P$ for fixed $R$.}
    \label{fig:1D_results}
\end{figure*}
Since $R$ is a proxy for the success rate of a random experiment, we investigate how $R$ changes with $M$ and $P$ for a system under various dynamics (i.e., pressure fields or mode shapes) and problem dimensionalities. We provide a heuristic for $R(M,P)$ in an idealized 1D problem,  then we investigate more complex and general 2D fields. 
\subsection{Ideal 1D Problem}\label{res:ideal-1d}
When using the first $M$ 1D resonant modes to move $P$ particles, the heuristic $R(M,P) = 1 - \frac{P}{M}$ is accurate within $3$ percentage-points (p.p.), as illustrated by the black solid (heuristic) and dashed (data) lines in Fig. \ref{fig:1D_results}. Note that the relationship between $R,M$ and $P$ shown in the figure's bottom plot implies that by keeping the ratio between the number of modes and particles constant results in constant $R$. Thus, if the number of particles is doubled, then the number of modes needs to be doubled to keep the controllability of the system constant. The equation was derived by exploiting the fact that $R(M,1)=1-\frac{1}{M}$ holds and generalizing the relation to multiple particles based on simulated data for $(M,P,R)$ described in \ref{subsec:r}. For more details, see Appendix \ref{appendix:r}.
\subsection{Unstructured Dynamics: Wendel's Theorem}\label{res:wendel}

When the system is at state $\boldsymbol{q}$, $\bm{F}_M(\boldsymbol{q})$ can be interpreted as $M$ directions around the origin of the state-space, as shown in Fig. \ref{fig:locally}. In the ideal 1D case, these directions are well structured, with dynamics given by Eq. \ref{eq:dotx}. In real experiments, these ideal shapes rarely appear and are commonly distorted or disturbed by noise and unmodeled perturbations, such as the presence of bubbles. As a consequence, idealized heuristics for $R$ might lose their predictive power in real settings. Thus, it is of interest to investigate how unstructured pressure fields affect $R$.

In order to tackle the ACP under unstructured dynamics, we invoke a classic result from Geometric Probability known as Wendel's theorem \cite{Wendel_1962}, formally described in Theorem \ref{theorem:wendel}. It provides a formula for the probability that $M$ points uniformly sampled from a unit-sphere in $\mathbb{R}^D$ centered at the origin lie on the same hemisphere, i.e., that their convex hull \textit{does not} contain the origin.
\begin{theorem}[\textbf{Wendel's Theorem}]
    \label{theorem:wendel}
     Let $M$ points be scattered at random on the surface of the unit sphere in an $D$-dimensional space. The probability that all points lie on some hemisphere is given by
    \begin{equation}
        \mathbb{P}_{M,{D}} = \frac{1}{2^{M-1}} \sum_{k=0}^{{D}-1} \binom{M-1}{k}.
    \end{equation}
\end{theorem}
If we take $D=\dim{\mathcal{S}^P}$  and each of the $M$ points to be a vector $f_m(\boldsymbol{q})$, then $\mathbb{P}_{M,D}$ is the complement of the probability that state $\bm{q}$ is locally controllable under unstructured dynamics. That is, if $W(M,D)$ is the probability that $\bm{q}$ is locally controllable, then
\begin{equation}\label{eq:wendel}
    W(M,D) = 1 - \mathbb{P}_{M,{D}} = 1- \frac{1}{2^{M-1}}\sum_{k=0}^{{D}-1} \binom{M-1}{k}.
\end{equation}

In a 2D problem, $W(M,2P)$ matches our definition for $R$ if the dynamics are random and i.i.d. Although, the assumption of i.i.d. phase velocity vectors is violated, as they are derived from a potential field, $W$ provides a good approximation for $R$ if the pressure fields contain a large degree of randomness or noise. Moreover, the shape of Eq. \ref{eq:wendel} is approximately linear in $D,M$, as shown at the bottom plot of Fig. \ref{fig:1D_results}, and, as a consequence of the Binomial Theorem, $W \approx 0.5$ when $D \approx M/2$.
\begin{figure}[H]
    \centering
    \begin{tabular}{c}
        \includegraphics[width=\linewidth]{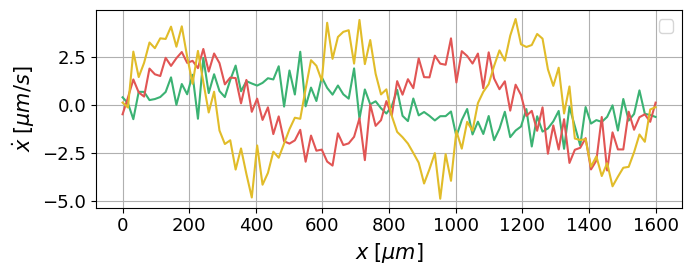} \\
        $\alpha=0.5$, low noise \\ \\
        \includegraphics[width=\linewidth]{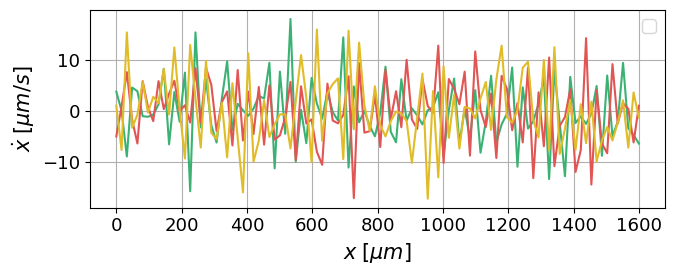} \\
        $\alpha=5$, high noise
    \end{tabular}
    \caption{Particle velocity in a 1D problem under the effect of system's first (green), second (red) and third (yellow) resonance modes with an artificially generated noise with levels $\alpha=0.5$ (top) and $\alpha=5$ (bottom).}
    \label{fig:noise1d}
\end{figure}
\begin{figure*}
    \centering
    \includegraphics[width=\textwidth]{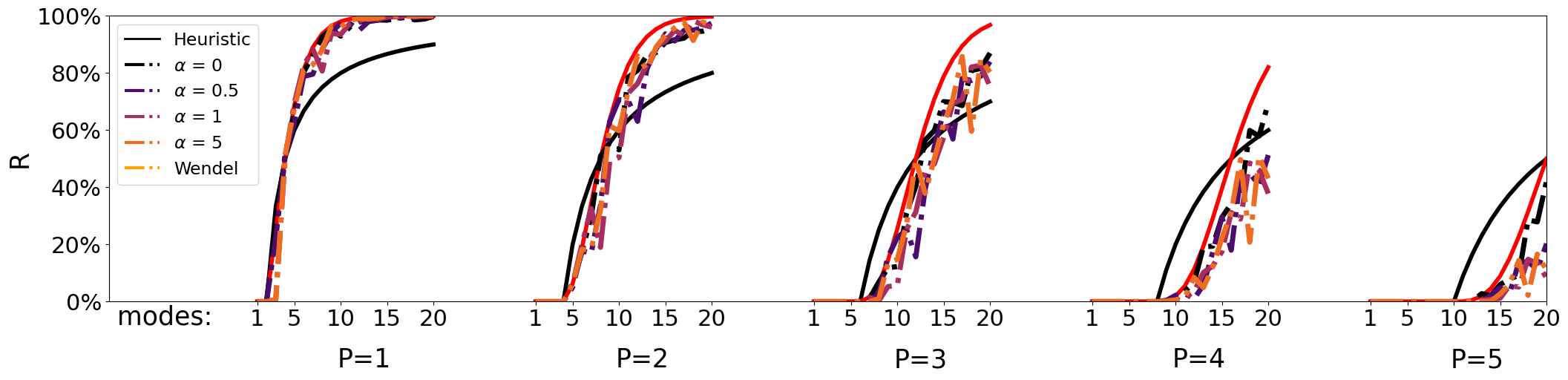}
\includegraphics[width=\textwidth]{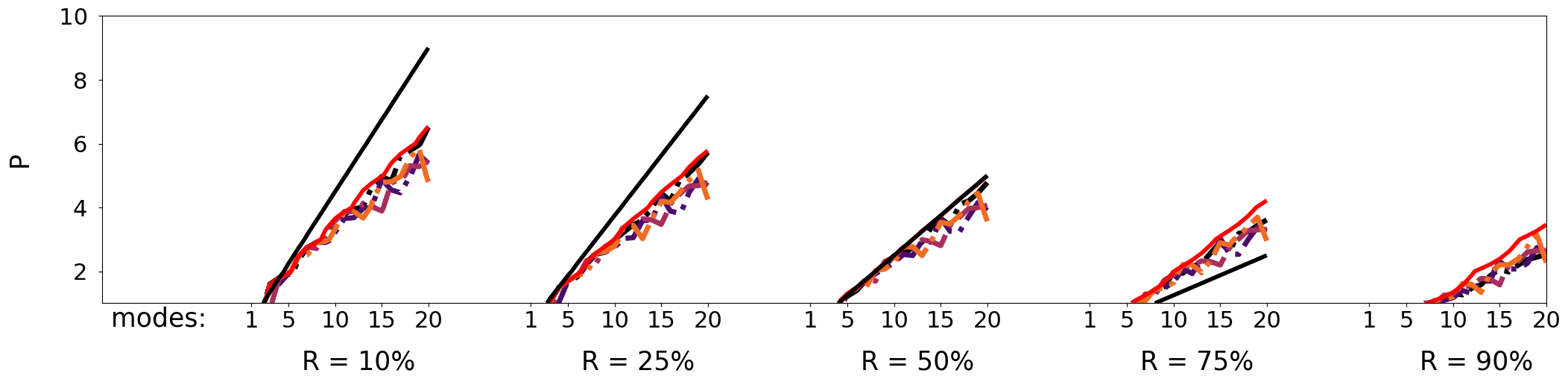}
    \caption{The local controllability ratio $R$ is numerically approximated using Monte Carlo sampling with varying number of particles  $P$, modes $M$ and noise levels $\alpha$ in the 2D problem. Additionally, the heuristic $R = 1-P/M$ and the Wendel approximation are shown, respectively, as black and red solid lines. The top plot emphasizes the effect of $P$ and $M$ on $R$, while the bottom one focuses on the linear relationship between $M$ and $P$ for fixed $R$.}
    \label{fig:2D_results}
\end{figure*}
\subsection{Non-ideal 1D Problem} \label{res:non-ideal1d}
In \ref{res:ideal-1d}, we made two assumptions: that the pressure fields are derived from ideal 1D standing waves, and that the $M$ available modes are provided in an orderly fashion, e.g., if $M=3$ modes are available, then the controller can actuate exactly the system's first, second, and third resonant modes. Here, we show that when either of these assumptions is relaxed, $R(M,P)$ starts behaving like $W(M,P)$.

First, smooth noise is added to the idealized potential to model a perturbation induced breakdown of structure, while keeping the modes ordered. The noise is generated by partial Fourier series with random coefficients, and a hyperparameter $\alpha$ controls the noise level of the system, as depicted in Fig. \ref{fig:noise1d}. Appendix  \ref{appendix:noise} describes in detail how the noise is generated. As a second experiment, we account for the assumption of orderly modes by gathering new data where we sample the $M$ modes uniformly from $\{1,2,...,100\}$. The results of both experiments are shown in Fig. \ref{fig:1D_results}, and they convey that these randomization procedures make the contour levels of $R$ transition from the idealized case towards $W(M,P)$.
\subsection{Non-Ideal 2D Problem}\label{res:non-ideal2d}

Although the 1D problem lends itself to a great toy example for theoretical investigations, real-life applications generally require at least two spatial dimensions per particle. Thus, it is of great importance to investigate how $M,P$ affect these 2D system's controllability under the realistic assumptions of additive noise and unordered modes.

Unlike their 1D counterparts, 2D modes are defined by two mode numbers and, thus, we sample $M$ from $\{m_x^{(1)}, m_x^{(2)}, \dots,m_x^{(M_{\text{max}})} \} \times \{m_y^{(1)}, m_y^{(2)}, \dots,m_y^{(M_{\text{max}})} \}$. Similar to the non-ideal 1D problem, we use partial 2D Fourier series with random coefficients to simulate additive noise into the system.

We simulated data for the 2D problem using $M_{\text{max}} = 100$ and the same noise levels used in the 1D problem. The results, shown in Fig. \ref{fig:2D_results}, show that, while adjusting the 1D heuristic to 2D (solid black line) does not provide as good results as before, ${W}(2P,M)$ (red line) is still an accurate metric for $R(M,P)$. Most importantly, at each controllability level, $M$ and $P$ still present a linear relationship.

In all cases, Wendel's formula Eq. \ref{res:wendel} provides a precise description of the $50\%$ controllability density.
\subsection{Numerical Verification: Pick and Place}\label{res:experiments}
One of our results is that to keep the level of controllability constant, the ratio of $M$ and $P$ needs to be held constant too. In this section, this statement is validated by numerical experiments, where we show that if a task can be achieved for a low number of particles and modes, it can also be achieved for a large number of particles as long as the ratio between the number of particles and the number of modes is held constant.

\begin{figure}[h]
    \centering
    \begin{tabular}{cc}
        \includegraphics[width=0.49\linewidth]{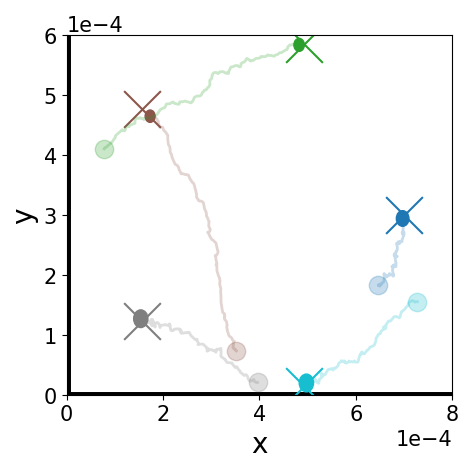} &   
        \includegraphics[width=0.49\linewidth]{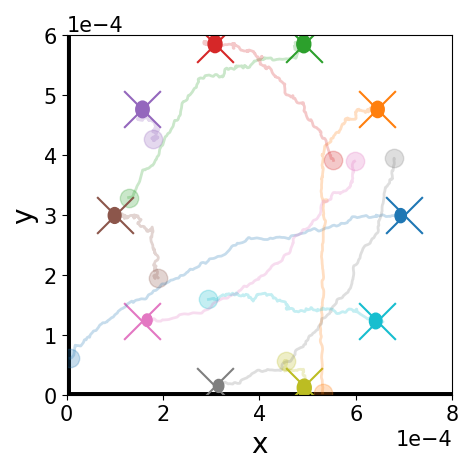} \\
        (a) $P=5, M=50$ & (b) $P=10, M=100$ \\
        \includegraphics[width=0.49\linewidth]{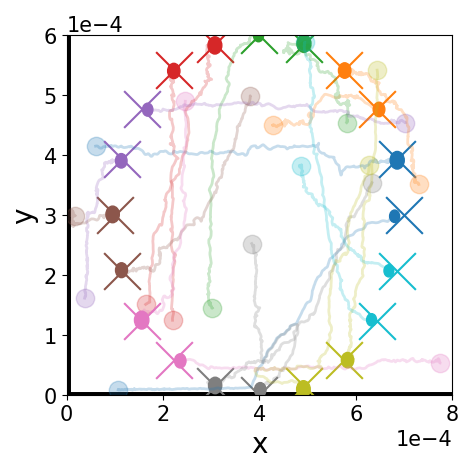} &   
        \includegraphics[width=0.49\linewidth]{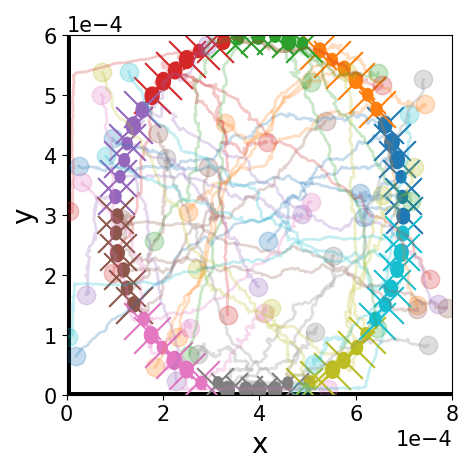} \\
        (c) $P=20, M=200$ & (d) $P=60, M=600$ \\
    \end{tabular}
    \caption{Illustration of four experiments where $P$ particles are taken from random initial positions towards target positions around a circle inside the device using $M$ resonant modes. Target positions are shown as crosses, while particle trajectories are shown as faded lines. The experiments were all successfully, and they lasted \textbf{(a)} 445, \textbf{(b)} 502, \textbf{(c)} 1121 and \textbf{(d)} 1902 iterations.}
    \label{fig:experiment}
\end{figure}

As done so far, we limit our particles to the first quarter of the device. Therefore, all initial positions and targets will be picked from this part of the device, as shown in Fig. \ref{fig:experiment}. The experiment starts by sampling a random position for each particle. Then, each particle's target is chosen from equidistant points on a circle inscribed in the first quadrant. Finally, PILOT takes each particle to its target position. We kept the ratio of $M$ to $P$ constant at $M/P=10$, while setting $P=5, 10, 20, 60$. The experiment was deemed successful if it was able to reach tolerance $\theta = 10$\textmu m within 5000 iterations.

The results, depicted in Fig. \ref{fig:experiment}, show that all four experiments were deemed successful, and they all finished within 2000 iterations, although the number of iterations increased considerably with the number of particles.

\section{Conclusion}

We analyzed the feasibility of a multimodal acoustophoresis control tasks theoretically and numerically. We defined the local controllability ratio, denoted as $R$, and showed that it is strongly correlated to the success rate $S$ of a random acoustophoresis control problem.  $R$ is purely a property of the pressure field geometry it can be computed much more efficiently than $S$. As $R$ condenses the total system performance to a single number, it can be used for experimental and numerical optimization of multimodal manipulation systems. Under the assumption of random and i.i.d. phase velocity vectors, $R$ can be computed directly using Wendel's theorem. Although the phase velocity factors are not necessarily i.i.d. in practice, the predictions based on the theorem fit the numerical experiments well as soon as a source of randomness (noise or randomly selected modes) is added.  In both 1D and 2D scenarios under noise-free and noisy dynamics, our results suggest that there is a linear relationship between $M$ and $P$ for a fixed controllability level, and that, in particular, $M \geq 2DP$ guarantees $R \geq 50\%$, where $D$ is the dimensionality of the problem. 

In order to verify these findings, we used a perfect-information local optimization algorithm, denoted PILOT, to control a system with 5, 10, 20 and 60 particles while keeping $M=10P$. All experiments were successful, and we were able to control 60 particles with 600 resonance modes.

\section{Acknowledgments}

D.W. has received funding from the European Research Council (ERC) under the European Union’s Horizon 2020 research and innovation programme under grant agreement No 834142 (ScalableControl) and from the Swedish Research Council (No. 2022-04041).
T.B. is supported by the SONOCRAFT project, funded by the European Innovation and Research Council (GA: 101187842), the Swedish Research Council (No. 2022-04041), and the Crafoordska Stiftelsen (No. 20240891).

\appendix

\section{$R(M,1)=1-\frac{1}{M}$}\label{appendix:r}


Since our problem is 1D and we have a single particle, then $ \boldsymbol{q} = x \in (0,L)$ for device length $L$. For resonant modes $m=1,2,...M$, it is possible to show that a state is locally controllable if, and only if, it is in the interval between the first and last stable equilibria of mode $M$, which are given by $\frac{L}{2M}$ and $L - \frac{L}{2M}$, respectively. Therefore, $R(M,1)$ will be the ratio between this interval and $L$, i.e., $R(M,1) = |\left( \frac{L}{2M}, L - \frac{L}{2M} \right)|/L = 1 - \frac{1}{M}$.

Generalizing the formula for $R(M,1)$ to multiple particles is a challenging analytical task, but one that can be done heuristically with high accuracy by exploiting numerical data from \ref{subsec:r}. The heuristic $\hat{R}(M,P)= 1-\frac{P}{M}$ was found to match the data very confidently, with an error bound of approximately 3 pp. 

Here, we show that in 1D systems with ideal dynamics and a single particle, $R=1-\frac{1}{M}$ as long as the controller is able to induce the first, second, ..., $M^{\text{th}}$ resonant modes. The proof is done with a unit-length device, given that its result is scalable for any $L$-long device. 

From the definition of local controllability, a state $x \in (0,1)$ is locally controllable if $\exists  m_1,m_2$ such that $f_{m_1}(x)f_{m_2}(x)<0$, i.e., if a particle is at $x$, there is at least one mode that can make it move forward and another that can make it move backward. Now, we will show that with $M$ available modes, the interval $I_M =\left(\frac{1}{2M}, \frac{2M-1}{2M} \right)$ is the set of all locally controllable states, which means that $R(M,1) = |I_M| = \frac{2M-1}{2M} - \frac{1}{2M} = 1 - \frac{1}{M}$.

At mode $m$, $f_m(x)$ will have an odd number of zeros at positions $\{\frac{1}{2m}, \frac{2}{2m}, ...,\frac{2m-1}{2m}\}$, with $f_m(x)$ alternating signs after each zero, starting with $(+)$ for $x<\frac{1}{2m}$ and ending on $(-)$ for $x > \frac{2m-1}{2m}$. Note that all variations in sign occur inside interval $I_m =(\frac{1}{2m}, \frac{2m-1}{2m})$, and that $I_1 \subset I_2\subset \dots \subset I_M$. This means that if we are able to use modes $m=1,2,...,M$, then the sign of $f_m(x)$ is constant in $x \notin I_M$ for all $m \leq M$, i.e., $x \text{ locally controllable } \implies x \in I_M$.

When we analyze two subsequent modes, we have $I_{m+1} = \left( \frac{1}{2(m+1)}, \frac{1}{2m} \right) \cup I_m \cup \left(\frac{2m-1}{2m}, \frac{2(m+1)-1}{2(m+1)} \right)$. Because $\frac{1}{2(m+1)} < \frac{1}{2m} \leq \frac{2}{2(m+1)}$, we have that for $x \in  \left( \frac{1}{2(m+1)}, \frac{1}{2m} \right)$, $f_{m+1}(x)<0$ and $f_m(x)>0$, therefore $\left( \frac{1}{2(m+1)}, \frac{1}{2m} \right)$ is locally controllable. A similar logic can be used to find that $\left( \frac{1}{2(m+1)}, \frac{1}{2m} \right)$ is also locally controllable, which means that so is $I_{m+1} - I_m$. The the union of locally controllable regions is locally controllable, so, because $I_{m+1} - I_m$ is always locally controllable, then we have that $I_M = \left( \cup_{m=2}^M \left( I_m - I_{m-1} \right) \right) \cup I_1$. But $I_1 = (\frac{1}{2}, \frac{1}{2}) = \emptyset$, which means that $I_M = \cup_{m=2}^M \left( I_m - I_{m-1} \right)$ is the union of locally controllable sets and is then locally controllable. So, $x\in I_M \implies x \text{ locally controllable}$. 

Therefore, $x \text{ locally controllable } \iff x \in I_M$ and $R(M,1) = |I_M| = 1 - \frac{1}{M}$.

\section{Additive Noise}\label{appendix:noise}

For each mode, its noise is generated by a partial real-valued Fourier series whose coefficients are sampled from a uniform distribution from -1 to 1. Formally, for a single mode $m$, we sample
\begin{equation}
    a_{n_x,n_y},b_{n_x,n_y}, c_{n_x,n_y}, d_{n_x,n_y} \, \overset{\text{i.i.d.}}{\sim} \,\mathcal{U}(-1,1),
\end{equation}
for $n_x,n_y=1,2,\dots,N_F$,
where $N_F$ denotes the length of the truncated series for each of the dimensions. From \cite{fourier}, the random partial Fourier series at position $(x,y)$ in the cavity is given by 
\begin{align*}
    \mathcal{F}(x,y) = \sum_{n_x=0}^{N_F} \sum_{n_y=0}^{N_F} \left(
        a_{n_x,n_y}
        \sin \left( 
            2\pi n_x \frac{x}{W} 
            \right) 
        \cos \left( 
            2\pi n_y \frac{y}{H} 
            \right) 
        \right.
        +  \\ \left.
        b_{n_x,n_y}
        \sin \left( 
            2\pi n_x \frac{x}{W} 
            \right) 
        \sin \left( 
            2\pi n_y \frac{y}{H} 
            \right) \right.
        + \\ \left.
        c_{n_x,n_y}
        \cos \left( 
            2\pi n_x \frac{x}{W} 
            \right) 
        \cos \left( 
            2\pi n_y \frac{y}{H} 
            \right) \right.
        + \\ \left.
        d_{n_x,n_y}
        \cos \left( 
            2\pi n_x \frac{x}{W} 
            \right) 
        \sin \left( 
            2\pi n_y \frac{y}{H} 
            \right) \right)
\end{align*}
When noise is used, the product of $\mathcal{F}$ and noise level $\alpha$ is added to $U$, creating the new potential
\begin{equation}
    \bar{U} = U + \alpha \mathcal{F},
\end{equation}
which, from Eq. \eqref{eq:force}, can be used to find the noisy acoustic-radiation force
\begin{equation}
    \bar{\bm{F}}_a = - \bm{\nabla} \bar{U} \implies 
    \bar{\bm{F}}_a = - \bm{\nabla}  U - \alpha \bm{\nabla}\mathcal{F}.
\end{equation}
Eq. \eqref{eq:dotx} can then be used with $\bar{\bm{F}}_a$ as the acoustic radiation force to describe how $\dot{\bm{x}}_p$ changes with the induced mode under noise level $\alpha$. This process is repeated for all $M$ modes, creating $M$ different random partial Fourier series, which are added to the mode's Gorkov potential.

\newpage

\bibliography{ref}

\end{document}